\documentclass[twocolumn,superscriptaddress]{revtex4-1}
\usepackage{amsmath}
\usepackage{amsfonts}
\usepackage{color}
\usepackage{graphicx}
\usepackage{subfigure}
\usepackage{blindtext}
\usepackage{hyperref}
\usepackage{natbib}

\newcommand{\ie}{\mbox{i.e.}~}
\newcommand{\eg}{\mbox{e.g.}~}
\newcommand{\cf}{\mbox{c.f.}~}
\newcommand{\comment}[1]{}

\newcommand{\rot}{\mathbf{\nabla \times}\ } 
\renewcommand{\div}{\mathrm{div}\ }

\newcommand{\imag}{\mathrm{i}}
\newcommand{\total}{\mathrm{d}}

\newcommand{\WSC}{\mathrm{WSC}}
\newcommand{\BZ}{\mathrm{BZ}}
\newcommand{\Lagrange}{\mathrm{L}}
\newcommand{\Energy}{\mathrm{U}}
\newcommand{\Green}{\mathcal{G}}
\newcommand{\Isofreq}{\mathcal{I}}

\newcommand{\LDOS}{\rho_\mathrm{t}}
\newcommand{\DOS}{\rho_\mathrm{t}}

\newcommand{\eps}{\varepsilon}

\newcommand{\eqnref}[1]{Eq.~\eqref{#1}}

\newcommand{\myvec}{\mathbf}

\begin{document}

\title{Modal expansions in periodic photonic systems with material loss and dispersion}

\author{Christian~Wolff}
\address{Center for Nano Optics, University of Southern Denmark, Campusvej 55, DK-5230~Odense~M, Denmark}
\email{cwo@mci.sdu.dk}
\author{Kurt~Busch}
\address{Max--Born--Institut, Max--Born--Stra{\ss}e 2A, D-12489 Berlin, Germany}
\address{Humboldt--Universit\"at zu Berlin, Institut f\"ur Physik, AG Theoretische Optik \& Photonik, Newtonstra{\ss}e 15, D-12489 Berlin, Germany }
\author{N.~Asger~Mortensen}
\address{Center for Nano Optics, University of Southern Denmark, Campusvej 55, DK-5230~Odense~M, Denmark}
\address{Danish Institute for Advanced Study, University of Southern Denmark, Campusvej 55, DK-5230~Odense~M, Denmark}

\begin{abstract}
  We study bandstructure properties of periodic optical systems composed of 
  lossy and intrinsically dispersive materials.
  To this end, we develop an analytical framework based on adjoint modes of a 
  lossy periodic electromagnetic system and show how the problem of linearly
  dependent eigenmodes in the presence of material dispersion can be overcome.
  We then formulate expressions for the bandstructure derivative 
  $(\partial \omega) / (\partial \myvec k)$ (complex group velocity) and 
  the local and total density of transverse optical states.
  Our exact expressions hold for 3D periodic arrays of materials with arbitrary 
  dispersion properties and in general need to be evaluated numerically.
  They can be generalized to systems with two, one or no directions of 
  periodicity provided the fields are localized along non-periodic directions.
  Possible applications are photonic crystals, metamaterials, metasurfaces
  composed of highly dispersive materials such as metals or lossless photonic 
  crystals, metamaterials or metasurfaces strongly coupled to resonant 
  perturbations such as quantum dots or excitons in 2D materials. 
  For illustration purposes, we analytically evaluate our expressions for some 
  simple systems consisting of lossless dielectrics with one sharp Lorentzian 
  material resonance added.
  By combining several Lorentz poles, this provides an avenue to perturbatively
  treat quite general material loss bands in photonic crystals.
\end{abstract}

\maketitle

\section{Introduction}

The concept of bandstructures (BS) is at the very heart of condensed-matter physics~\cite{Kohn:1999}, where the wave dispersion relations of electrons, phonons etc. are discussed almost entirely in terms of real-valued energies and wave vectors, forming also a natural starting point for important concepts such as the \emph{group velocity} and the \emph{density-of-states} (DOS). In these systems with lossless and energy-independent Hamiltonians, group velocity and DOS are intimately linked. It should be stressed, that this relationship between the bandstructure derivative (group velocity) and the imaginary part of the Green tensor (DOS) is not trivial and in fact no longer holds for lossy and explicitly energy-dependent Hamiltonians, as we found. Obviously, the lossless theory developed in solid-state theory was a fruitful starting point also for common accounts for light waves in periodic dielectric structures~\cite{Busch:2007}, \ie photonic crystals (PhC). As such, the real bandstructures (\ie real-valued frequencies as functions of real-valued wave vectors) of photonic crystals composed of loss-less and non-dispersive materials have been studied in detail~\cite{Joannopoulos:2008,Johnson:2001}, as well as additional investigations of loss-less, but dispersive problems~\cite{Raman:2010}.
This includes the retrieval of effective permittivities and permeabilities in the long-wavelength limit~\cite{Movchan:2003,Menzel:2008,Alu:2011}, the study of stop band and complete band gaps~\cite{Ho:1990,Yablonovitch:1991,Krauss:1996,Blanco:2000}, and the investigation of singular points, such as Dirac~\cite{Haldane:2008} or Weyl points~\cite{Lu:2013}, van-Hove singularities and the density-of-states~\cite{McPhedran:2004}, and band edges in the context of slow light~\cite{Baba:2008,Notomi:2010}.
In many cases, the focus was on the modification of thermal, spontaneous or stimulated emission in the aforementioned unusual dispersion regimes~\cite{Dowling:1994,Lodahl:2004,Lu:2014,Hoeppe:2012,Florescu:2007,Schuler:2009}. Here, the key quantity is the group velocity 
$\myvec v = \partial \omega / (\partial \myvec k)$, whose inverse (the group slowness) is proportional to the photonic DOS for lossless, dispersionless periodic systems. The focus on lossless systems can be motivated by the applications' emphasis on transparent optical materials and the obvious requirement of low loss for lasers or single-photon sources. 

On the other hand, even inherently weak absorption will be enhanced in the slow-light regime~\cite{Pedersen:2008,Grgic:2012}, thus emphasizing the importance of including optical absorption in photonic bandstructure theory.
Furthermore, optical absorption is crucial for the study of thermal emission and moreover, some studies on enhancing Purcell factors at band edges used PhC infiltrated with many narrow-band emitters, which are bound to modify the
optical properties of the structure. The same holds true to even greater extent for lasing in photonic crystals.
Thus, a systematic investigation of the impact of narrow-band spectral perturbations to PhCs seems justified.

From a mathematical point of view, loss and in particular dispersive material 
response pose the problem that the solutions to the wave equation no longer
form a Hilbert space or even a basis of the function space they span.
The reason underlying this is the fact that the macroscopic Maxwell equations 
do not cover the complex physical processes and internal degrees of freedom 
that result in the dispersive permittivity of a non-idealized dielectric.
One remedy to this is to include these degrees of freedom as auxiliary
polarization fields to the master equation, leading to states that form a 
Hilbert space and whose time-evolution is described by a unitary 
operator~\cite{Tip:1998}.
This comes at the expense of a vast (effectively infinite) number of 
auxiliary differential equations and while without any doubt it led to very
fundamental insight into the physics of lossy electromagnetic structures,
it is a little unwieldy for numerical evaluation.
In contrast, we accept the mathematical limitations and derive expressions
for key quantities such as the band structure derivative and the DOS from
only the electromagnetic fields using the notion of adjoint states and
adjoint operators.
The present study is not restricted to classic photonic crystals. We explicitly allow for an already lossy and dispersive initial structure, which especially includes metallic metamaterials and plasmonic lattices~\cite{Soukoulis:2011,Abajo:2007}. It furthermore applies to any bulk material through an empty lattice approach and to any system that can be obtained as the limit of a uniformly converging sequence of periodic systems, especially including planar waveguides and states bound to two-dimensional (2D) materials or metal-dielectric interfaces supporting surface-plasmon polaritons (SPP)~\cite{Torma:2014}.
We justify this claim in Sec~\ref{sec:conclusions}.

The remainder of this paper is structured as follows.
In the Sec.~\ref{sec:problem}, we precisely state the problem covered in this
paper and in Sec.~\ref{sec:assumptions}, we state our assumptions and 
approximations.
This is followed by our main section~\ref{sec:general_prop}, in which we 
lay out the basic mathematical tools for the treatment of periodic systems with
lossy and dispersive constituents.
Specifically, we introduce a family of bilinear forms to replace the common 
scalar product, we explicitly state the adjoints of the Maxwell operator and 
its solutions, we derive the band structure derivative (complex group velocity)
and finally the transversal DOS.
In Sec.~\ref{sec:examples}, we demonstrate the relevance of our finding by 
analytically discussing three simple examples: a homogeneous material, a 
perturbation theory for non-degenerate bands in arbitrary periodic structures
and the case of an isotropic band edge with a Lorentzian material resonance 
added as a model system for the self-limiting of strong Purcell factor 
enhancement by the emitters themselves, as one might find in PhC-based lasers.
The paper ends with a discussion and conclusions and is followed by three
appendices.

\section{The problem of bandstructures in dispersive systems}
\label{sec:problem}

\begin{figure}
  \includegraphics[width=\columnwidth]{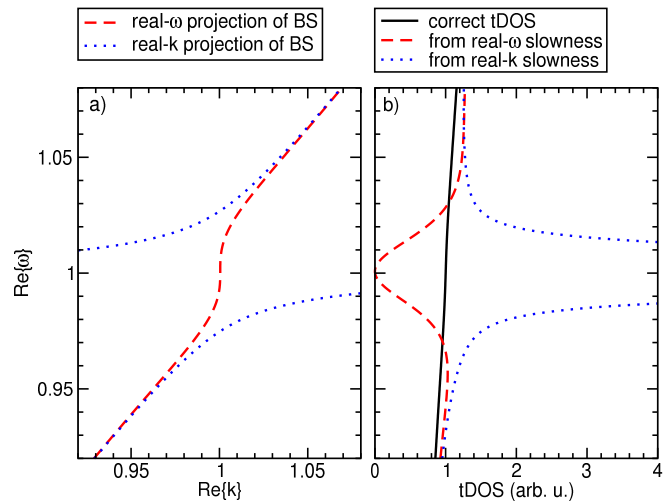}
  \caption{
    (color online) 
    Example for the difficulty of deriving the DOS from common bandstructure
    plots.
    The system is a gas of strong Lorentzian oscillators in vacuum.
    The frequency bandstructure is really a complex function of complex
    wave number.
    Panel a) shows two typical (and equally correct) projections: purely real 
    frequency vs real part of the (complex) wave number and real part of the 
    (complex) frequency associated with purely real wave numbers.
    Panel b) shows the correct DOS (derived from the Green tensor) and 
    incorrect results found by naively computing a ``DOS'' from the apparent 
    group velocities of either curve in panel a).
    The example is taken from the discussion in Sec.~\ref{sec:bulk_example}.
  }
  \label{fig:intro_example}
\end{figure}

The theme of this paper is the effect of material dispersion and loss on the 
propagation of optical pulses and the emission dynamics of excited two-level 
systems in periodic media.
As a prototypical source of loss and dispersion, we often explicitly study 
Lorentzian resonances (\eg caused by color centers or quantum dots).
The central quantities for pulse propagation and photon emission are the group 
velocity and the density-of-states, respectively.
While both are very well defined and intuitive in lossless structures, their
meaning can become slightly obscure in the presence of loss and dispersion.

This is, because -- despite the common focus on just real-valued wave vectors 
and real frequencies -- the dispersion relation (\ie the frequency function) is 
in fact a multi-valued complex function of complex wave 
vectors~\cite{Reuter:2017}.
More precisely, the frequency bands are the zeros of an operator that depends 
analytically on $\myvec k$ and are therefore analytic functions of the wave 
vector away from branch points, which appear for example at the boundaries and 
the center of the first Brillouin zone.
The branch points connect the individual frequency sheets, which are the 
continuations of the energy bands into the complex $\myvec k$-space.
This fact is well known within the solid-state theory community and has been 
used \eg to analyze the localization properties of electronic and photonic 
Wannier functions~\cite{Heine:1963,He:2001,Busch:2011} (see also other 
references within Ref.~\cite{Reuter:2017}).
Since such a manifold is already hard to visualize for a single direction of
periodicity (complex $\omega$ and a scalar complex $k$ would require a 
4-dimensional plot), it is instructive to restrict oneself to contours where 
either $\omega$ or $\myvec k$ is purely real-valued.
We refer to them as ``real-$\omega$'' contours and ``real-$\myvec k$'' contours,
respectively and they often play a special role in the analysis of the band 
structure, especially for Fourier transforms.
The reason for this is that a Fourier transform in time corresponds to an 
integral over real $\omega$ and a (lattice-)Fourier transform in space 
corresponds to an integral over the first real Brillouin zone.
In the absence of loss, this is easily dealt with, because all frequencies
associated with real $\myvec k$ are real, hence the real-$k$ contour
is part of the real-$\omega$ contour; Fourier transforms with respect to space
are compatible with transforms in time.
However, even for lossless problems, the real-$\omega$ contour has substantial
contributions with complex $\myvec k$.
In electronic crystal theory, they are known as ``Heine's lines of real 
energy''~\cite{Heine:1963}. 
However, these purely evanescent states usually don't contribute to the bulk 
properties (\eg the bulk Green tensor) of a lossless system and therefore 
are usually ignored.

With the inclusion of loss, the picture changes significantly, because the
eigenstates in the real $\myvec k$-space now correspond to complex frequencies,
\ie, the contour of real $\omega$ no longer includes the real $\myvec k$-space.
Conversely, solutions with real-valued $\omega$ appear away from the real 
$\myvec k$-axis.
Further practical problems arise, because the eigenstates of a lossy system
are not orthogonal in the usual way and those of a dispersive system can be
even linearly dependent, hence not forming a basis (as we show in this paper).
Some researchers try to circumvent these problems by neglecting the imaginary 
part of either $\omega$ or $\myvec k$, \eg by evaluating the bandstructure in 
the real $\myvec k$ space, ignoring imaginary parts in order to treat this
``sanitized'' function $\omega(\myvec k)$ as a conventional (lossless) 
bandstructure, potentially re-introducing the loss perturbatively in a second 
step.
Such a treatment may work acceptably for very weak loss, but suffers from the
fundamental problem that ``sanitizing'' the bandstructure (\ie ignoring 
imaginary parts) can be done in at least two ways: either ignoring the imaginary
part of $\myvec k$ when solving for the real-$\omega$ contour. 
This can be justified by the fact that the DOS is a function of the (real) 
frequency.
Alternatively, one can ignore the imaginary part of $\omega$ when solving the
bandstructure in the real $\myvec k$-space, which can be equally justified,
because a real $\myvec k$ is the appropriate quantum number to label physical
states in a periodic system.
The probably more damning problem is that this approach of ``sanitizing'' the
complex bandstructure can produce substantially incorrect results.
We illustrate this by the simple case of a homogeneous isotropic material (the 
simplest possible periodic structure) with one Lorentzian material resonance,
which has been studied in depth about twenty years 
ago~\cite{Barnett:1996,Scheel:1999}.
In the example shown in Fig.~\ref{fig:intro_example} (see 
Sec.~\ref{sec:bulk_example} for further details), the ``sanitized'' real-$k$ BS 
shows two bands with \emph{vanishing} group velocity at the resonance 
frequency, where the ``sanitized'' real-$\omega$ BS only shows one band with 
\emph{infinite} group velocity at the resonance frequency.
While the ``DOS'' computed from the group velocity on the real-$\omega$ contour
predicts a substantial dip and potentially vanishing DOS right at the resonance,
the ``DOS'' computed from the group velocity on the real-$k$ contour predicts
a divergence around the same frequencies; the correct DOS (more precisely the
transversal DOS~\cite{Barnett:1996,Scheel:1999}, see next paragraph) however is 
nearly constant across the resonance in this example (it can diverge in extreme
cases).
This illustrates that a more detailed analysis is required to correctly predict
the DOS of lossy periodic systems.

Some key points of the theory of a homogeneous material with Lorentzian 
resonance~\cite{Barnett:1996} are that the emission dynamics of a quantum 
emitter in a lossy environment must be based on the imaginary part of the Green 
tensor, where the total density-of-states (DOS) comprises two contributions: 
a radiative transversal DOS (tDOS) derived from the Green tensor of modes 
that satisfy Maxwell's divergence conditions as well as a non-radiative 
longitudinal DOS (lDOS) due to evanescently excited longitudinal polarization 
fields.
This is an illustration of the fact that the divergence-free solutions of the 
electromagnetic wave equation do not form a complete basis~\cite{Sipe:2000}.
However, the influence of these quasi-static longitudinal modes is restricted
to an atomic length scale in lossy dielectrics and metals below the plasma
frequency.
Finally, the physically relevant DOS differs from the imaginary part of
the Green tensor by local field correction factors (derived \eg from
Clausius--Mossotti theory or through similar approaches~\cite{Barnett:1996}) 
to account for the atomic structure of physical materials.
While the material-dependence of the lDOS and the local field correction 
factors are usually restricted to an atomic length scale, the tDOS is 
determined by long-range (quasi-)propagating optical modes.
Therefore, substantial differences between the DOS for (lossy and dispersive)
homogeneous and periodic arrangements of such materials can be expected to
emanate from the tDOS, while local corrections based on the immediate host
material of the dipole can be expected to provide a decent approximation.
Therefore, in the following we restrict our analysis of the DOS to the tDOS 
without local field corrections.
It should be noted that this restriction becomes inadequate if one of the 
constituent materials supports long-range longitudinal modes, \eg bulk
plasmons.
In this case, the long-range corrections to the lDOS must be revised 
analogously to Sec.~\ref{sec:DOS}.

\section{Assumptions and notation}
\label{sec:assumptions}

We consider a three-dimensional (3D) periodic electromagnetic system (photonic crystal or metamaterial) composed of materials with dispersive, lossy and anisotropic but spatially local dielectric response, where any frequency-dependence is assumed to be smooth.
At one point, we must assume that the material properties also vary smoothly in space, although on an arbitrarily short length scale. In the following, we will use capital letters to emphasize physical fields in the time or frequency domain, such as full Bloch modes, while lower case letters denote the lattice-periodic parts of Bloch modes in the frequency domain. They are connected via the Bloch--Floquet theorem. For the electric field, this is:
\begin{align}
  \myvec E_{n \myvec k}(\myvec r, t) = \myvec e_{n \myvec k}(\myvec r, \omega)
  \exp(\imag \myvec k \myvec r - \imag \omega t),
\end{align}
where mathematically both $\omega$ and $\myvec k$ can be complex-valued. 
We emphasize that it is only through further physical considerations that we may confine $\omega$ and $\myvec k$ in the complex plane, \eg assuming $\omega$ real as appropriate for a discussion of the photonic states that can be excited by a spectrally well-defined continuous wave from a narrow-linewidth laser. 

Within the perturbation theory of Sec.~\ref{sec:bs_derivative}, we will assume a Bloch mode with wave vector $\myvec k$ that is perturbed in the Cartesian $z$-direction $\hat z = (0,0,1)^T$ without loss of generality, because the relative orientation of the system and the Cartesian axes is arbitrary.

Within the derivation of the DOS in Sec.~\ref{sec:DOS}, we will expand the Green tensor in an auxiliary basis constructed for each physical frequency $\omega$. All quantities related to this auxiliary basis are highlighted by an additional subscript $\omega$. Finally, we assume reciprocity for the derivation of the Green tensor.

\section{General properties of lossy dispersive periodic structures}
\label{sec:general_prop}

In this section, we will first introduce a set of mathematical
tools to study lossy yet nondispersive periodic media.
This includes the definition of a family of bilinear forms to replace the 
common scalar product for physical states and the introduction of adjoint 
fields and wave operators with respect to a chosen bilinear form.
We then show how to generalize this for the case for lossy and dispersive
problems, after which we derive the (complex) bandstructure derivative of an 
arbitrary lossy, dispersive periodic system.
We close this section with a derivation of the transverse electromagnetic 
density-of-states inside a general lossy, dispersive periodic system.

\subsection{Non-dispersive eigenvalue problem: adjoints and orthogonality}
\label{sec:nondispersive_evp}

We start our discussion with Maxwell's curl equations in frequency domain,
\begin{align}
  \mu^{-1} \rot \myvec E = & \imag \omega \myvec H,
  \\
  \eps^{-1} \rot \myvec H = & -\imag \omega \myvec E,
\end{align}
where $\eps$ and $\mu$ are complex-valued, tensorial functions of $\myvec r$, 
which we assume not to depend on frequency for the moment.
We also note that we exclude solutions that violate Maxwell's divergence 
equations, which results in an incomplete function space~\cite{Sipe:2000} and
ultimately to the distinction between transversal and longitudinal DOS.
The corresponding equations for the lattice periodic parts are:
\begin{align}
  \mu^{-1} (\nabla + \imag \myvec k) \times \myvec e = & \imag \omega \myvec h,
  \\
  \eps^{-1} (\nabla + \imag \myvec k) \times \myvec h = & -\imag \omega \myvec e.
\end{align}
To ease our later notation, this can be formally simplified by introducing an operator $\mathcal{L}(\myvec k)$ and state vectors $\Psi$ or $\psi$ for physical or 
lattice-periodic states, respectively:
\begin{align}
  \Psi = & \left( \begin{array}{c}
      \myvec H \\ \myvec E
  \end{array} \right),
  \quad\quad\quad
  \psi = \left( \begin{array}{c}
      \myvec h \\ \myvec e
  \end{array} \right),
  \\
  \mathcal{L}_0 = & \left( \begin{array}{cc}
      0 & -\imag \mu^{-1} \nabla \times \\
      \imag \eps^{-1} \nabla \times & 0
  \end{array} \right),
  \\
  \mathcal{L}(\myvec k) = & \left( \begin{array}{cc}
      0 & -\mu^{-1} (\imag \nabla - \myvec k) \times \\
      \eps^{-1} (\imag \nabla - \myvec k) \times & 0
  \end{array} \right).
\end{align}
With this, Maxwell's equations reduce to:
\begin{align}
  \mathcal{L}_0 \Psi = & \omega \Psi, &
  \mathcal{L}(\myvec k) \psi = & \omega \psi,
  \label{eqn:eigenproblem_fd}
\end{align}
for stationary eigenstates and their lattice-periodic parts, respectively.
The latter is a regular eigenvalue problem (EVP) for the angular frequency 
$\omega$ and an implicit eigenvalue problem for the wave vector $\myvec k$ 
(it can be transformed into a regular EVP for one component of $\myvec k$).

In the lossy case, the operator $\mathcal{L}(\myvec k)$ is not self-adjoint 
with respect to any standard scalar product and as a consequence the 
eigenmodes are not orthogonal in a conventional sense.
We decide to allow for maximal flexibility and thus introduce a family of 
bilinear forms, which maps a pair $\Psi_{1,2}$ of frequency-domain states 
(not necessarily eigenstates) to a generally 
complex scalar:
\begin{align}
  (\Psi_1, \Psi_2) = \lim_{N\rightarrow \infty} 
  \frac{1}{V_N} \int_{V_N} \total^3 r \ \Psi_1(\myvec r) 
  \mathcal{W}(\myvec r) \Psi_2(\myvec r),
\end{align}
where the integration is carried out over volumes $V_N$ composed of $N$ 
Wigner--Seitz cells (WSC).
This is a straight-forward generalization of how a scalar product and 
orthogonality between eigenstates is introduced in (lossless) solid state 
theory~\cite{Haken:1976}.
The various bilinear forms are determined by the choice of the weight function 
$\mathcal{W}(\myvec r)$ that has the periodicity of the lattice and must 
evaluate to a full-rank $6\times 6$ matrix everywhere.
Using the Bloch--Floquet theorem, this can be decomposed:
\begin{align}
  (\Psi_1, \Psi_2) = & \overbrace{
  \lim_{N\rightarrow \infty} \frac{1}{N} \sum_{\myvec R \in V_N} 
  \exp[\imag (\myvec k_1 + \myvec k_2) \cdot \myvec R)] }^{
    \mathcal{S}_{\myvec k_1, \myvec k_2}
  }
  \nonumber
  \\
  & \quad 
  \times \frac{1}{V_\WSC} \int_\WSC \total^3 r \ 
  \psi_1(\myvec r) \mathcal{W}(\myvec r) \psi_2(\myvec r),
  \intertext{with}
  \mathcal{S}_{\myvec k_1, \myvec k_2} = & \left\{
  \begin{array}{lcl}
    1 & , &\text{for} \,\, \myvec k_1 + \myvec k_2 = 0 \\
    \infty & ,& \text{for} \,\, \Im\{\myvec k_1 + \myvec k_2\} \neq 0 \\
    0 & ,& \text{otherwise}
  \end{array}
  \right. .
\end{align}
As a result, Bloch states (but not lattice-periodic wave functions) associated 
with opposite imaginary part of $\myvec k$ are orthogonal to each other and we 
can restrict ourselves to the WSC-integral:
\begin{align}
  (\psi_1, \psi_2) = & \int_\WSC \total^3 r \ 
  \psi_1(\myvec r) \mathcal{W}(\myvec r) \psi_2(\myvec r).
\end{align}
As a crucial next step, we can now formally introduce the adjoint state $\psi^\ddagger$ and the adjoint operator $\mathcal{L}^\ddagger(\myvec k)$, which 
satisfy with respect to a given choice of $\mathcal{W}$:
\begin{align}
  \big(\mathcal{L}^\ddagger(\myvec k) \psi_1^\ddagger, \psi_2\big) =
  \big(\psi_1^\ddagger, \mathcal{L}(\myvec k) \psi_2\big),
\end{align}
for an arbitrary pair of lattice-periodic parts and for any wave vector 
$\myvec k$.
For a pair of appropriately normalized, non-degenerate eigenstates, they 
therefore satisfy:
\begin{align}
  (\psi_n^\ddagger, \psi_{m}) = \delta_{nm},
\end{align}
where $\delta_{nm}$ is the Kronecker delta. 
Note that the exact form of the adjoint operator and the adjoint modes depend 
on the choice of bilinear form.

The most natural (but not necessarily always best suited as demonstrated 
throughout this paper) choice is the ``energy form'' defined by
\begin{align}
  \mathcal{W}^{(\Energy)}(\myvec r) = & 
  \frac{1}{2} \left( \begin{array}{cc}
  \mu(\myvec r) & 0 \\
  0 & \eps(\myvec r) 
  \end{array} \right).
  \label{eqn:energy_form_def}
\end{align}
It is most natural, because it is the straight-forward generalization of the
standard scalar product in lossless, dispersionless dielectric structures.
It can be shown (see Appendix~\ref{appx:adjoint}) that for this bilinear 
form the adjoint operator is
\begin{align}
  \mathcal{L}^\ddagger(\myvec k) = \left( \begin{array}{cc}
      0 & \mu^{-1} (\imag \nabla + \myvec k) \times \\
      -\eps^{-1} (\imag \nabla + \myvec k) \times & 0
  \end{array} \right),
  \label{eqn:adjoint_operator_energy_nondispersive}
\end{align}
and consequently the adjoint of any state $\Psi(\myvec r, \omega, \myvec k)$ with angular frequency $\omega$ and wave vector $\myvec k$ is given as:
\begin{align}
  \Psi^\ddagger(\myvec r, \omega, \myvec k) = 
  \Psi(\myvec r, -\omega, -\myvec k),
  \label{eqn:adjoint_mode_connection_nondispersive}
\end{align}
modulo an arbitrary factor.
Thus, we can conclude for adjoint modes with respect to the energy form:
\begin{align}
  (\Psi_{n \myvec k}^\ddagger, \Psi_{m \myvec q}) = &
  \delta_{nm} \delta(\myvec q - \myvec k),
\end{align}
with $\delta(\ldots)$ being the Dirac delta function, while $\myvec k$ and $\myvec q$ are two complex wave vectors with equal 
imaginary part, \ie $\Im\{\myvec q - \myvec k\} = 0$.
In other words: for a non-dispersive system, all eigenstates with the same 
$\Im\{\myvec k\}$ and different eigenfrequencies are orthogonal to each other.


\subsection{Mathematical tools for dispersive problems}
\label{sec:dispersive_evp}

Next, we consider the case where the material response functions $\eps$ and
$\mu$ are dispersive.
As a result, both the Maxwell operator $\mathcal{L}(\myvec k)$ and the weight 
function $\mathcal{W}$ depend on frequency.
The stationary solutions satisfy the equation:
\begin{align}
  \mathcal{L}(\myvec k, \omega) \psi_{n \myvec k} = \omega \psi_{n \myvec k}.
  \label{eqn:dispersive_evp_def}
\end{align}
This is now an implicit eigenvalue problem for both the angular frequency 
$\omega$ and the wave vector $\myvec k$.
While it can still be transformed into a regular EVP for one component of 
$\myvec k$, this is no longer possible with respect to $\omega$, because 
Eq.~\eqref{eqn:dispersive_evp_def} is in general non-linear in $\omega$.

One major problem with dispersive systems is that the set of physical solutions
for a given wave vector is no longer orthogonal in the sense of the previous
section.
Even worse, a resonance in $\eps$ or $\mu$ in general also introduces additional 
photonic bands and makes the set of physical eigenstates linearly dependent
(\cf Appendix~\ref{appx:multivalue}).
As a result, the eigenstates no longer form a basis of the function space they
span and for example the familiar expression of the Green tensor in terms of
eigenstate-projectors divided by energy poles --- \ie of the form
$\int \total k \ | k \rangle \langle k | / (\omega - \omega_k)$ ---
is no longer applicable.
A third, more subtle but equally prohibitive problem is the incompleteness of 
the physical eigenstates introduced by Maxwell's divergence 
condition~\cite{Sipe:2000}. 
In a non-dispersive system, the physical eigenstates associated with different 
eigenfrequencies satisfy the same divergence condition and therefore form an
appropriate expansion basis for a physical field distribution (\eg the 
transverse Green tensor) at any frequency.
In contrast, the eigenstates of the dispersive problem satisfy \emph{different}
divergence conditions and therefore they cannot be used to represent a 
physical field distribution at any specific frequency.

This can be only solved by removing the nonlinearity in the eigenvalue, \ie by
fixing $\omega$ in one way or another and separating it from the eigenvalue.
The first way is to use $\omega$ and some components of $\myvec k$ as
parameters and study the eigenvalue problem with the remaining component of
$\myvec k$ as the eigenvalue.
This approach has advantages for numerical bandstructure 
calculations~\cite{Hermann:2008,Tserkezis:2009}, where it often referred to as 
the on-shell approach.

The way we will use in the following is to regard both $\omega$ and $\myvec k$ 
as fixed parameters and introduce a new variable $\lambda$ as a frequency-like 
eigenvalue:
\begin{align}
  \mathcal{L}(\myvec k, \omega) \phi_{n \myvec k \omega} = 
  \lambda_{n \myvec k \omega} \phi_{n \myvec k \omega}.
  \label{eqn:auxiliary_problem}
\end{align}
This recovers the physical states whenever 
$\omega = \lambda_{n \myvec k \omega}$, which is also sometimes used to compute
bandstructure relations for dispersive system by starting with an initial
guess for $\omega$, setting $\omega = \lambda$ and repeating this until
self-consistency is achieved~\cite{Hermann:2008}.
Much more importantly in our context however is the fact that the eigenfunctions
of Eq.~\eqref{eqn:auxiliary_problem} are linearly independent (yes only 
orthogonal in the sense of Sec.~\ref{eqn:dispersive_evp_def}) and thereby
provide a convenient expansion basis for mathematical objects associated with 
a given $\omega$, especially the Green tensor and consequently the DOS 
(see Sec.~\ref{sec:DOS}).
Since every physical frequency $\omega$ requires a different auxiliary problem,
we designate quantities related to Eq.~\eqref{eqn:auxiliary_problem} with a
subscript $\omega$, whereas quantities related to the original physical, 
dispersive problem~\eqref{eqn:eigenproblem_fd} lack this subscript.
Eq.~\eqref{eqn:auxiliary_problem} provides a convenient expansion basis for 
stationary problems at frequency $\omega$, because the eigenstates form a 
complete basis.
At the same time, we also fix $\omega$ in the weight function $\mathcal{W}$.
In this sense, we find based on the ``energy form'' for arbitrarily dispersive
material response that can be  represented in time domain as a real-valued 
memory kernel, \ie for material functions that satisfy the relations
$\eps(-\omega) = \eps(\omega)$ and $\mu(-\omega) = \mu(\omega)$:
\begin{align}
  \mathcal{L}^\ddagger(\myvec k, \omega) 
  = &
  -\mathcal{L}(-\myvec k, -\omega),
  \\
  \Psi^\ddagger(\myvec r, \omega, \myvec k) = &
  \Psi(\myvec r, -\omega, -\myvec k),
  \label{eqn:adjoint_mode_connection_dispersive}
\end{align}
again modulo an arbitrary factor.

\subsection{Band structure derivative}
\label{sec:bs_derivative}

Next, we derive the expression for the derivative of $\omega$ with respect to 
$\myvec k$.
To this end, we perturb these two quantities:
\begin{align}
  \omega' = \omega + \Delta \omega, \quad\quad
  \myvec k' = \myvec k + \kappa \hat z,
  \label{eqn:perturbations}
\end{align}
with in general complex scalars $\Delta \omega$ and $\kappa$.
As mentioned earlier, the assumption that $\myvec k$ is perturbed in the 
$z$-direction does not lead to any loss of generality.
It is natural to generalize the notion of the group velocity to the complex 
bandstructure derivative $(\partial \omega) / (\partial \myvec k)$ in a lossy 
system and while some authors~\cite{Chen:2010} are hesitant about the physical 
meaning of the imaginary part of a complex group velocity, 
others~\cite{Gerasik:2010} have interpreted it in the context of a loss rate.
This discussion is beyond the scope of this paper and in order to avoid 
confusion, we will refer to the term $(\partial \omega) / (\partial \myvec k)$ 
just as the bandstructure derivative for the remainder of this paper.

Since the weight function $\mathcal{W}$ may be dispersive, the bilinear form 
in general depends on the eigenvalue $\omega$:
\begin{align}
  ( \psi_1, \psi_2 )_\omega = &
  \int_{\WSC} \total^3 r \ \psi_1 \mathcal{W}(\myvec r, \omega) \psi_2.
\end{align}
We now perturb the operator in \eqnref{eqn:eigenproblem_fd}:
\begin{align}
  \big(\mathcal{L}(\myvec k) + \kappa \mathcal{P}_z\big) \psi = \omega' \psi,
  \label{eqn:operator_perturbation_vgr}
\end{align}
with the perturbation operator:
\begin{align}
  \mathcal{P}_z = \left(
    \begin{array}{cc}
      0 & \mu^{-1} \hat z \times \\
      -\eps^{-1} \hat z \times & 0
  \end{array} \right).
\end{align}
Next, we project onto the adjoint mode $\psi^\ddagger$:
\begin{align}
  \big(\psi^\ddagger, 
  [\mathcal{L}(\myvec k) + \kappa \mathcal{P}_z] \psi\big)_{\omega'}
  = & \omega' \big(\psi^\ddagger, \psi\big)_{\omega'}.
\end{align}
Note that the bilinear forms and all operators must be evaluated at the 
perturbed frequency.
We now use the identity 
$\big( \psi^\ddagger, \mathcal{L}(\myvec k) \psi \big)_\omega = 
\omega \big( \psi^\ddagger, \psi \big)_\omega$, and we introduce Taylor 
expansions of the form
\begin{align}
  \mathcal{W}(\omega + \Delta \omega) =
  \mathcal{W}(\omega) 
  + \Delta \omega \frac{\partial \mathcal{W}}{\partial \omega} \Big|_\omega
  + \mathcal{O}(\Delta \omega^2).
\end{align}
for $\mathcal{W}$, $\mathcal{L}(\myvec k)$ and $\mathcal{P}_z$.
In the integral definition of the bilinear form, 
we assume that $\mathcal{O}(\Delta \omega^2)$ is at least of order 
$\mathcal{O}(\kappa^2)$, because $\Delta \omega$ and $\kappa$ are related via 
the complex group velocity, which we assume to be bounded:
\begin{align}
  & 
  \Delta \omega 
  \int_\WSC \total^3 r \ \psi^\ddagger \ 
  \frac{\partial\{\mathcal{W} [\omega - \mathcal{L}(\myvec k)]\}}{\partial \omega}
  \ \psi 
  \\
  = &
  \kappa \int_\WSC \total^3 r \ \psi^\ddagger 
  \mathcal{W} \mathcal{P}_z \psi 
  + \mathcal{O}(\kappa^2),
  \nonumber
\end{align}
where the symbols $\mathcal{W}$, $\mathcal{L}$ and $\mathcal{P}_z$ are
evaluated at the unperturbed frequency $\omega$.
From this, we can now easily find the projected BS derivative along the 
$z$-direction:
\begin{align}
  v_z = & \lim_{\kappa \rightarrow 0} \frac{\Delta \omega}{\kappa}
  = \frac{\mathcal{F}}{\mathcal{N}};
  \label{eqn:vgr_general}
  \\
  \mathcal{F} = & \int_\WSC \total^3 r \ \psi^\ddagger \mathcal{W}(\omega) \mathcal{P} \psi;
  \label{eqn:F_general}
  \\
  \mathcal{N} = & \int_\WSC \total^3 r \ \psi^\ddagger 
  \frac{\partial \{\mathcal{W} [\omega - \imag \mathcal{L}(\myvec k)]\}}{\partial \omega} 
  \psi.
  \label{eqn:N_general}
\end{align}
This is a very general form of the BS derivative for complex $\myvec k$ and
complex $\omega$.

The quantities $\mathcal{N}$ and $\mathcal{F}$ as well as the definition of the
adjoint modes depend on the choice of $\mathcal{W}$, but the expression
$\mathcal{F}/\mathcal{N}$ is always the BS derivative.
Therefore, we may conclude that whenever it is possible to interpret 
$\mathcal{N}$ as a conserved physical density per WSC, then $\mathcal{F}$ must 
be the associated flux through the WSC.
In the case of the ``energy form'', the product $\mathcal{W} \mathcal{L}$ does
not depend on $\omega$ and the denominator becomes
\begin{align}
  \mathcal{N}^{(\Energy)} = \frac{1}{2} \int_\WSC \total^3 r \ 
    \myvec e^\ddagger \frac{\partial (\eps \omega)}{\partial \omega} \myvec e 
    + \myvec h^\ddagger \frac{\partial (\mu \omega)}{\partial \omega} \myvec h 
  = \mathcal{U},
\end{align}
which is reminiscent of the modal energy of a (lossy and dispersive) Bloch mode.
Conversely, $\mathcal{F}^{(\Energy)}$ resembles the Poynting flux integrated 
over the unit cell:
\begin{align}
  \mathcal{F}^{(\Energy)} = 
  \frac{1}{2} \int_\WSC \total^3 r \ \hat z \cdot 
  (\myvec e \times \myvec h^\ddagger + \myvec e^\ddagger \times \myvec h).
\end{align}
The explicit form of the BS derivative with respect to the ``energy form'' is:
\begin{align}
  v_z = & \frac{
  \int_\WSC \total^3 r \ \hat z \cdot (
  \myvec e^\ddagger \times \myvec h
  + \myvec e \times \myvec h^\ddagger 
  )
  }{
  \int_\WSC \total^3 r \ 
    \myvec e^\ddagger \frac{\partial (\eps \omega)}{\partial \omega} \myvec e 
    + \myvec h^\ddagger \frac{\partial (\mu \omega)}{\partial \omega} \myvec h 
  }.
  \label{eqn:vgr_energy}
\end{align}

As we show in Appendix~\ref{appx:adjoint}, the adjoint modes with respect to
the ``energy form'' are given as:
\begin{align}
  \psi^\ddagger(\myvec r, \myvec k, \omega) 
  = C \psi(\myvec r, -\myvec k, -\omega),
\end{align}
with some constant $C$.
In the absence of loss (and only then), this reduces to the familiar 
relationship
$ \psi^\ddagger(\myvec r, \myvec k, \omega)
= \psi^\ast(\myvec r, \myvec k, \omega)$.
Another possible choice for the weight function is
\begin{align}
  \mathcal{W}^{(\Lagrange)} = 
  \frac{1}{2} \left( \begin{array}{cc}
  -\mu(\myvec r, \omega) & 0 \\
  0 & \eps(\myvec r, \omega) 
  \end{array} \right).
  \label{eqn:lagrangian_form_def}
\end{align}
This induces the ``Lagrangian form'' (we use picked name, because the integrand 
resembles the Lagrangian density, while the integrand of the ``energy form'' 
Eq.~\eqref{eqn:energy_form_def} resembles the electromagnetic energy density).
It leads to adjoint modes of the form
$\psi^\ddagger(\myvec r, \myvec k, \omega) 
= \psi(\myvec r, -\myvec k, \omega)$, \ie in lossless systems
$\psi^\ddagger(\myvec r, \myvec k, \omega)
= \psi^\ast(\myvec r, \myvec k, -\omega)$.
This is the definition employed in Ref.~\onlinecite{Chen:2010}.
Thus, it is no surprise that the ``Lagrangian form'' reproduces that expression
for the BS derivative:
\begin{align}
  v_z = & \frac{
  \int_\WSC \total^3 r \ \hat z \cdot (
  \myvec e^\ddagger \times \myvec h
  - \myvec e \times \myvec h^\ddagger 
  )
  }{
  \int_\WSC \total^3 r \ 
    \myvec e^\ddagger \frac{\partial (\eps \omega)}{\partial \omega} \myvec e 
    - \myvec h^\ddagger \frac{\partial (\mu \omega)}{\partial \omega} \myvec h 
  }.
  \label{eqn:vgr_lagrangian}
\end{align}

\subsection{Transversal densities of states}
\label{sec:DOS}

\begin{figure*}
  \includegraphics[width=\textwidth]{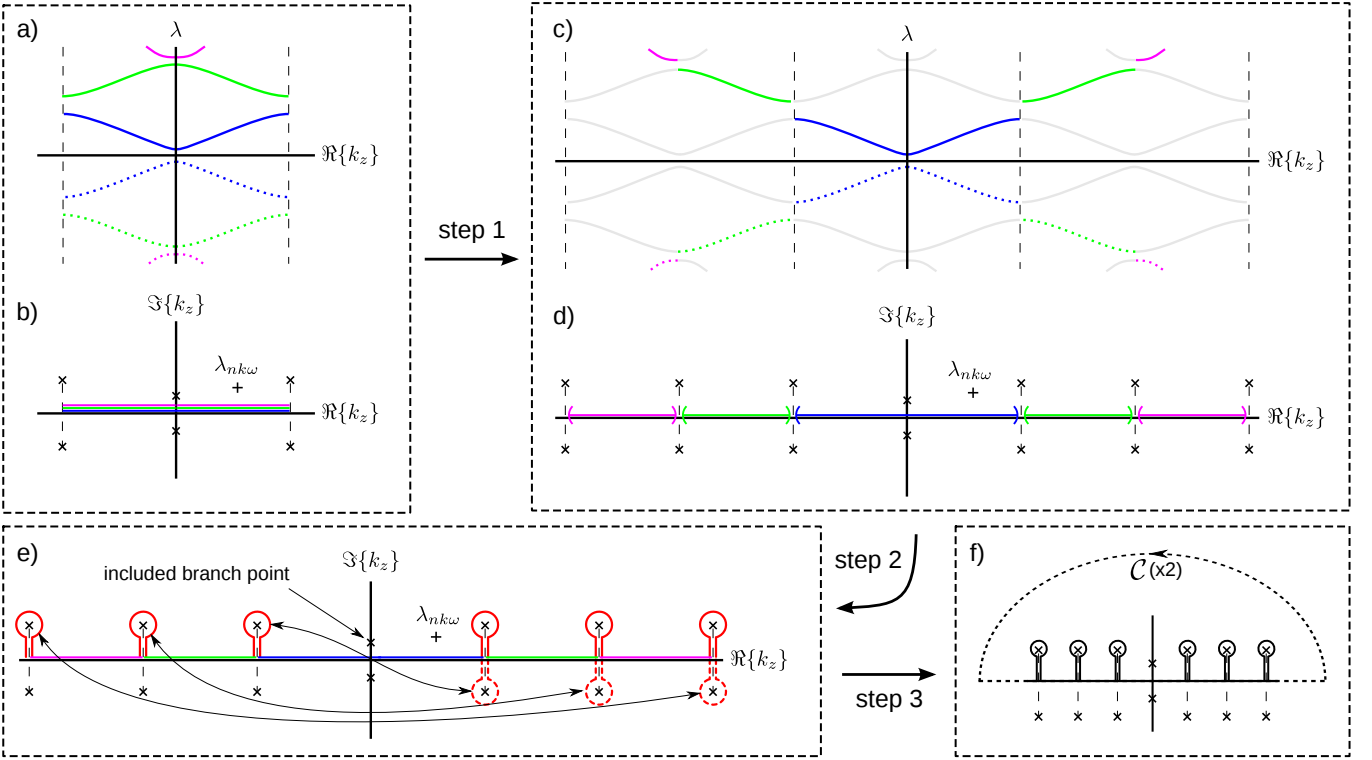}
  \caption{
    (color online)
    Sketch of the process of transforming the $\myvec k$-space integration for
    the case of a one-dimensional (1D) system stacked along the $\hat z$-axis
    with an arbitrarily small $k_x \neq 0$.
    The top left box shows the conventional integration procedure, where the
    different bands are distinguished by colors.
    Panel a: conventional bandstructure for a loss-less system.
    Panel b: outline of the reduced zone scheme integration path in the complex
    $k_z$-plane; all bands are integrated over the first BZ. 
    Each stop band corresponds to a pair of branch points (indicated as `x' 
    symbols ) at the BZ boundary connected by a branch cut (thin dashed line).
    The resonance pole of the Green function is indicated by a `+' symbol and
    lies away from the $k_z$-real axis for lossy systems. \\
    The top right box shows the transformation to two integrals (one solid, 
    one dotted line) in an extended zone scheme, \ie integration over the $i$th 
    band is shifted to the $i$th BZ; the individual integral segments remain 
    disconnected. 
    Panel c: Conventional dispersion diagram for a loss-less system.
    Each integral covers eigenvalues with opposite sign for positive and 
    negative $k_z$ to match the symmetry of adjoint modes 
    [see Eq.~\eqref{eqn:adjoint_mode_connection_dispersive}].
    Panel d: Extended zone scheme integration path in the complex plane.\\
    The bottom left box (panel e) shows the integration contour after 
    connecting the individual integral segments around the stop band branch 
    points except the one at $k_z = 0$.
    In the limit $\myvec r' \rightarrow \myvec r$, the integrand is symmetric
    under $(\myvec k, \lambda) \rightarrow (-\myvec k, \lambda)$ and the 
    hairpins are equivalent to a set of closed integral loops, each of which 
    vanishes.
    The complex plane above this integration contour includes exactly one
    branch point at $\Re\{k_z\} = 0$, \ie closing the contour around the top
    brings us from the positive $\lambda$-bands to negative bands and vice 
    versa.
    Integrating twice along the contour $\mathcal{C}$ shown in panel f provides
    a closed integral to which the residue theorem can be applied.
    See the main text and Appendix~\ref{appx:integration} for further details.
  }
  \label{fig:integration}
\end{figure*}

The transversal local density-of-states (tLDOS) $\LDOS(\omega, \myvec r)$ is 
fundamentally linked~\cite{Sheng:2006} to the the electromagnetic Green tensor
$\Green(\omega, \myvec r, \myvec r')$ evaluated for $\myvec r = \myvec r'$:
\begin{align}
  \LDOS(\omega, \myvec r) = 
  -\frac{2\omega}{\pi c^2} \Im\{\Green(\omega, \myvec r, \myvec r)\}.
\end{align}
Often, the Green tensor is expressed in terms of the eigenstates of the
physical system.
This becomes less straight-forward in a dispersive medium, because the physical 
eigenstates no longer form a basis.
Therefore, for any given frequency $\omega$, we use the linearly independent 
eigenfunctions derived from the auxiliary EVPs
\begin{align}
  \mathcal{L}_0(\omega) \Phi_{n \myvec k \omega} = &
  \lambda_{n \myvec k \omega} \Phi_{n \myvec k \omega},
  \\
  \mathcal{L}(\myvec k, \omega) \phi_{n \myvec k \omega} = &
  \lambda_{n \myvec k \omega} \phi_{n \myvec k \omega}
\end{align}
as an $\omega$-specific expansion basis.
This is meant merely as a mathematical tool to obtain a result that ultimately
does not contain any reference to the auxiliary EVP.
We can now find the Green tensor of the non-dispersive auxiliary problem:
\begin{align}
  \mathcal{L}_0(\omega) \Green_\omega(\lambda, \myvec r, \myvec r') =
  \delta(\myvec r - \myvec r'),
\end{align}
in terms of the basis $\Phi$, where the expansion coefficients are given as
the projection of the Dirac-source onto the adjoint modes $\Phi^\ddagger$.
We choose the adjoint modes to be normalized 
$(\Phi_{n\myvec k\omega}^\ddagger, \Phi_{m\myvec q\omega}) = \delta_{nm} 
\delta(\myvec k - \myvec q)$
with respect to the ``energy form''.
The resulting Green tensor is:
\begin{align}
  & \Green_\omega(\lambda, \myvec r, \myvec r') \\
  \nonumber
  & = 
  \sum_n \int_\BZ \total^3 k \ \frac{
    [\Phi_{n\myvec k \omega}(\myvec r) \otimes
    \Phi^\ddagger_{n\myvec k \omega}(\myvec r')]
  \mathcal{W}(\omega, \myvec r')}{\lambda - \lambda_{n \myvec k \omega}}.
  \label{eqn:green_expanded_100}
\end{align}
The integral covers all real wave vectors inside the first Brillouin zone and
the sum runs over all bands with both positive and negative energy.
Note that the denominator in general remains finite, because the real frequency
eigenvalues occur at $\myvec k$-vectors with nonzero imaginary parts.
Therefore the familiar identity 
\mbox{
  $(\lambda - \lambda_{n \myvec k \omega})^{-1} = 
\imag \pi \delta(\lambda - \lambda_{n \myvec k \omega}) + 
\mathfrak{P} (\lambda - \lambda_{n \myvec k \omega})^{-1}$ }
cannot be used, where $\mathfrak{P}$ represented the Cauchy principal value.
This would not be of limited use anyway, because the numerator is in general
complex-valued, \ie the resulting tDOS would contain the unwieldy principal 
value integral.
Instead, we transform this integral (defined over the first Brillouin zone)
to a contour integral to which the residue theorem can be applied.

In the remainder of this section, we now assume that the Cartesian 
$\hat z$-axis is parallel to one of the reciprocal lattice vectors and we assume
that $k_x$ and $k_y$ remain real-valued while $k_z$ will be allowed to take
non-real values.
The sum of integrals over all bands in Eq.~\eqref{eqn:green_expanded_100} can 
be recast into four integrals each over an extended path in complex 
$\myvec k$-space: one path with positive and one with negative $\lambda$ for 
each of the two independent polarizations.
This is illustrated in Fig.~\ref{fig:integration} for the simplified case
of a single polarization in 1D (stacked along the $\hat z$-direction): 
the first band is integrated over the first BZ, the integral over the second 
band is moved to the second BZ and so on.
Then, separate integrals over the different BZs are connected by ``hairpin'' 
paths running around the upper branch points at the BZ boundaries in the 
complex plane.
The connecting hairpins to not change the value of the integral in the limit 
$\myvec r' \rightarrow \myvec r$ 
(\cf Appendix~\ref{appx:integration_hairpin}), which 
is the appropriate limit for the calculation of densities of states.
Since the numerator $\Phi_{n\myvec k \omega}(\myvec r) \otimes 
\Phi^\ddagger_{n\myvec k \omega}(\myvec r')$ of 
Eq.~\eqref{eqn:green_expanded_100} is the product of two Bloch functions, it
implicitly contains a factor 
$\exp[\imag \myvec k \cdot (\myvec r - \myvec r')]$.
Therefore, we can close the integration contour at infinity in the upper 
complex half-plane of $k_z$, if we decide to approach the limit 
$\myvec r' \rightarrow \myvec r$ such that $z' < z$, leading to the integration
path $\mathcal{C}$ shown in Fig.~\ref{fig:integration}e.
One slight complication is that the contour $\mathcal{C}$ encircles a branch 
point at $\Re\{k_z\} = 0$.
Traversing it once takes us from the branch of positive $\lambda$ to the 
branch of negative $\lambda$ and vice versa.
Thus, traversing the contour $\mathcal{C}$ twice results in an integration path
that allows to apply the residue theorem.
This description of the case of a single polarization in 1D can be readily
generalized to two polarizations in 3D (\cf Appendix~\ref{appx:integration_3d}).

Following the above argument, the collocal Green tensor elements can be 
represented by the residues:
\begin{align}
  & \Green_\omega(\lambda, \myvec r, \myvec r) 
  \\
  \nonumber
  = &
  2 \pi \imag \int_{\Isofreq_{z,\omega}(\lambda)} 
  \!  \!  \!  \!
  \total^2 k \ 
  \frac{\partial k_z}{\partial \lambda}
    [\Phi_{n\myvec k \omega}(\myvec r) \otimes
    \Phi^\ddagger_{n\myvec k \omega}(\myvec r)] 
    \mathcal{W}(\omega, \myvec r).
    \label{eqn:DOS_deriv_100}
\end{align}
Here, $\Isofreq_{z,\omega}(\lambda)$ is the contour in complex $\myvec k$-space 
on which the auxiliary problem~\eqref{eqn:auxiliary_problem} takes the 
eigenvalue $\lambda$ with the additional constraint that $k_x$ and $k_y$ remain
real numbers.
This and the exact meaning of our notation is explained more precisely in
Appendix~\ref{appx:integration_3d}.
Finally, the special case $\Isofreq_{z,\omega}(\omega) = \Isofreq_z(\omega)$ 
is the isofrequency contour of the physical, dispersive problem 
Eq.~\eqref{eqn:eigenproblem_fd}.

We can now replace the partial derivative in Eq.~\eqref{eqn:DOS_deriv_100}
with the corresponding expression we found in Sec.~\ref{sec:bs_derivative} to
obtains an expression that no longer contains any reference to the auxiliary 
eigenvalue problem~\eqref{eqn:auxiliary_problem} and is therefore suitable to
express the collocal Green tensor elements for the original, physical, 
dispersive system~\eqref{eqn:eigenproblem_fd}:
\begin{align}
  &\Green(\omega, \myvec r, \myvec r) 
  \\
  \nonumber
  = &
  2 \pi \imag \int_{\Isofreq_z(\omega)} 
  \!  \!  \!  \!
  \total^2 k \ 
  \frac{ 
    [\Psi_{n\myvec k}(\myvec r) \otimes
    \Psi^\ddagger_{n\myvec k}(\myvec r)] 
  \mathcal{W}(\omega, \myvec r)}{
    \int_\WSC \total^3 r \ \hat z \cdot (
    \myvec e^\ddagger_{n\myvec k} \times \myvec h_{n\myvec k} 
    + \myvec e_{n\myvec k} \times \myvec h^\ddagger_{n\myvec k} 
    )
  }.
\end{align}
From this, we find for the tLDOS tensor:
\begin{align}
  & \LDOS(\omega, \myvec r) = - \frac{4 \omega}{c^2}
  \\
  \nonumber
  & \times \int_{\Isofreq_z(\omega)} 
  \!\!\!\!\! \total^2 k \ 
  \Re\bigg\{ 
    \frac{ 
      [\Psi_{n\myvec k}(\myvec r) \otimes
      \Psi^\ddagger_{n\myvec k}(\myvec r)] 
    \mathcal{W}(\omega, \myvec r)}{
      \int_\WSC \total^3 r \ \hat z \cdot (
      \myvec e^\ddagger_{n\myvec k} \times \myvec h_{n\myvec k} 
      + \myvec e_{n\myvec k} \times \myvec h^\ddagger_{n\myvec k} 
      )
  } \bigg\},
\end{align}
The total tDOS is the trace of the tLDOS-tensor integrated over
the WSC, \ie we replace the numerator in the tLDOS with
$
  \int_\WSC \total^3 r \ 
  \Psi^\ddagger_{n\myvec k}(\myvec r)
  \mathcal{W}(\omega, \myvec r)
  \Psi_{n\myvec k}(\myvec r) = 1
  $.
We find for the total tDOS:
\begin{align}
  & \DOS(\omega) = \frac{4 \omega}{c^2}
  \label{eqn:DOS_final}
  \\
  \nonumber
  & \times 
  \int_{\Isofreq_z(\omega)} 
  \!\!\!\!\! \total^2 k \ 
  \Re\bigg\{ 
    \frac{
      \int_\WSC \total^3 r \ 
      \myvec e^\ddagger_{n\myvec k} \eps \myvec e_{n\myvec k} + \myvec h^\ddagger_{n\myvec k} \mu \myvec h_{n\myvec k} 
    }{
      \int_\WSC \total^3 r \ \hat z \cdot (
      \myvec e^\ddagger_{n\myvec k} \times \myvec h_{n\myvec k} 
      + \myvec e_{n\myvec k} \times \myvec h^\ddagger_{n\myvec k} 
      )
  } \bigg\}.
  \\
  & = \frac{4 \omega}{c^2}
  \int_{\Isofreq_z(\omega)} 
  \!\!\!\!\! \total^2 k \ 
  \Re\bigg\{ 
    \frac{ 
      (\psi^\ddagger_{n\myvec k}, \psi_{n\myvec k}) 
    }{
      (\psi^\ddagger_{n\myvec k}, \mathcal{P}_z \psi_{n\myvec k}) 
  } \bigg\}.
\end{align}
In the non-dispersive, but potentially lossy case, the integrand is the real 
part of the inverse BS derivative (the group slowness in lossless systems).
In the dispersive case, it differs from the latter by a factor
\begin{align*}
  \frac{
    \int_\WSC \total^3 r \ 
    \myvec e^\ddagger \eps \myvec e + \myvec h^\ddagger \mu \myvec h 
  }{
    \int_\WSC \total^3 r \ 
    \myvec e^\ddagger \frac{\partial (\eps \omega)}{\partial \omega} \myvec e 
    + \myvec h^\ddagger \frac{\partial (\mu \omega)}{\partial \omega} \myvec h 
  }.
\end{align*}

\section{Band structures with material resonances}
\label{sec:examples}

Optical dispersion due to material resonances deforms the optical dispersion in 
a characteristic way.
In particular, Lorentzian resonances can lead to the emergence of additional 
optical bands (see also Appendix~\ref{appx:multivalue}), 
which are separated by an anti crossing near the resonance frequency.
This is closely related to the appearance of mini-gaps and ``bandstructure 
bubbles'' that have been found in systems involving a strong material 
resonance near a photonic band edge~\cite{Hermann:2008}. 
In essence, this is equivalent to local band backbending.
Other examples include the effect of a plasma resonance, leading \eg to the
well known surface plasmon polaritons in half-space problems and back-folded 
photonic bands in periodic systems~\cite{Chen:2011}. 
Oftentimes, these problems are studied ignoring optical loss in order to avoid
dealing with non-real frequencies or wave vectors.
In this section, we qualitatively discuss the effect of material resonances on 
dispersion relations, specifically on the BS derivative and the density-of-states.
We study this using a narrow-band Lorentzian resonance as an example, but the
results qualitatively apply to plasmonic resonances, as well.

\subsection{Homogeneous isotropic material with single resonance pole}
\label{sec:bulk_example}

\begin{figure}
  \includegraphics[width=\columnwidth]{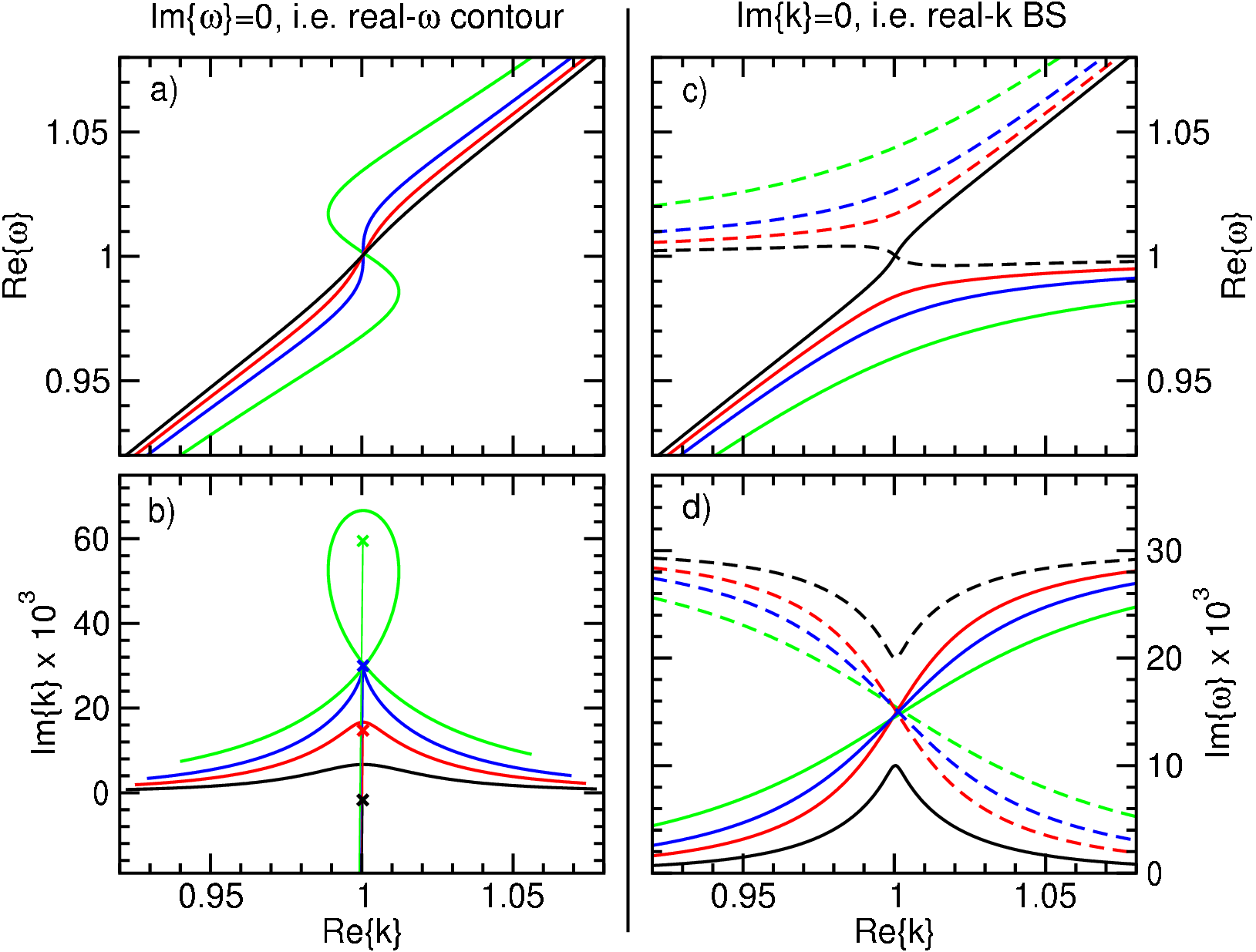}
  \caption{
    (color online)
    Complex bandstructure of an isotropic homogeneous medium with Lorentz 
    resonance ($\Omega = 1.0$, $\gamma = 0.03$).
    The line colors represent different oscillator strengths $A$:
    \mbox{$-0.4\times10^{-3}$} (black),
    \mbox{$-1.0\times10^{-3}$} (red),
    \mbox{$-1.8\times10^{-3}$} (blue),
    \mbox{$-4.0\times10^{-3}$} (green).
    The left column shows views of the \mbox{real-$\omega$} curve, \ie the 
    part of
    the path in complex \mbox{$k$-space} that is associated with real 
    frequencies.
    In contrast, the right column shows views of the \mbox{real-$k$} band 
    structure, \ie the complex frequencies associated with real wave numbers.
    Crosses and thin solid lines in panel~(b) indicate the resonance-related
    branch points and branch cuts.
    Solid and dashed lines indicate corresponding solutions in panels~(c) and 
    (d).
    See text for further discussion.
  }
  \label{fig:coupling_cases}
\end{figure}

The simplest periodic system is an isotropic bulk material.
Due to its simplicity, it can be solved analytically and the results show the
effect of resonant dispersion as clearly as possible. 
The bulk bandstructure of a homogeneous material with speed of light $c$ and
one additional resonant pole is given by the relation:
\begin{align}
  c^2 k^2 = \left( 1 + \frac{A}{\omega - \widetilde \Omega} \right) 
  \omega^2
  =
  \frac{\omega^2 (\omega - \widetilde \Omega + A)}{\omega - \widetilde \Omega}.
  \label{eqn:bulk_disp_relation}
\end{align}
where $k$ is the absolute value of the wave vector, $A$ is the oscillator
strength and $\widetilde \Omega = \Omega - \imag \gamma$ is its pole with 
undressed resonant frequency $\Omega$ and damping $\gamma$.
In the absence of dispersion, the function $\omega(k)$ is double-valued 
(one band includes the positive real $\omega$, the other the negative real
values).
The resonance increases the number of bands from two to three, so the function
$\omega(k)$ has three sheets.
The connecting branch points are the points of degeneracy, \ie where the
function
\begin{align}
  f(\omega) = \omega^3 - (\widetilde \Omega - A) \omega^2 - 
  c^2 k^2 \omega + c^2 k^2 \widetilde \Omega = 0
\end{align}
has a double zero.
These points can be found as the zeros of the derivative $f'(\omega)$:
\begin{align}
  \omega_\text{Br} = & \widetilde \Omega 
  - \frac{1}{4} \left(A \pm \sqrt{8\widetilde \Omega A + A^2}\right),
\end{align}
where the corresponding wave numbers follow from 
Eq.~\eqref{eqn:bulk_disp_relation}.
The position of these branch points and their connecting branch cut relative
to the real $k$-axis and the curve of real frequencies determines the topology
of the bandstructure.

\begin{figure}
  \includegraphics[width=\columnwidth]{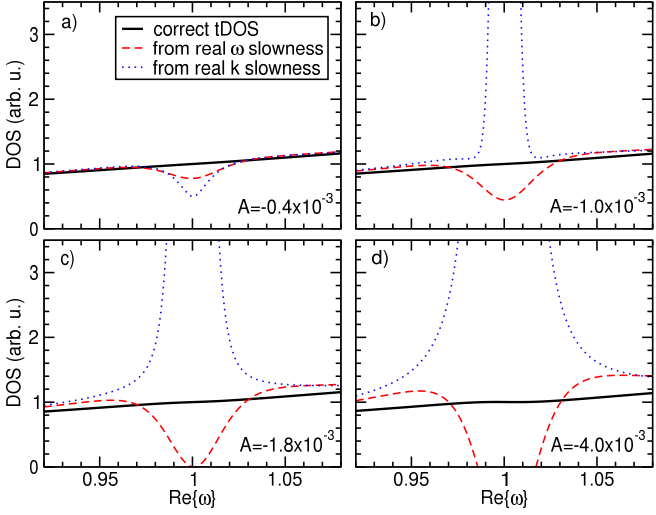}
  \caption{
    (color online)
    Comparison of the proper tDOS [Eq.~\eqref{eqn:DOS_final}, black solid line] 
    and incorrect expressions based on the group slowness derived from 
    ``sanitized'' real-$\omega$ (red dashed) and the real-$k$ (blue dotted) 
    contours for an isotropic bulk medium with a 
    Lorentz resonance ($\Omega = 1.0$, $\gamma = 0.03$) for four different 
    oscillator strengths (annotated in the graphs).
    See text for further discussion.
  }
  \label{fig:bulk_DOS_comparison}
\end{figure}

We identify four main cases as illustrated in Fig.~\ref{fig:coupling_cases}.
Weak oscillators (small parameter $A$) lead to a bandstructure that is 
qualitatively not different from the lossless case both on the real $k$-axis 
and on the real-$\omega$ curve (black curves).
As the oscillator strength increases, the topology on the real $k$-axis changes 
from a purely photonic band and an oscillator band separated in imaginary 
$\omega$-direction to a more conventional anticrossing exactly when the upper 
branch point (indicated by crosses in Fig.~\ref{fig:coupling_cases}) moves into 
the upper half plane (black and red lines in Fig.~\ref{fig:coupling_cases}).
Thus, the position of the upper branch point determines two regimes that might 
be identified as weak and strong coupling.
It should be stressed that the real-$\omega$ curve 
(solid lines in left column of Fig.~\ref{fig:coupling_cases}), which defines the
density-of-states, does not change qualitatively during this transition.
However, for very strong oscillators, the upper branch point can reach the
real-$\omega$ curve (blue curves in Fig.~\ref{fig:coupling_cases}).
This leads to a cusp, where the BS derivative is not defined (blue curves).
For even stronger oscillators, the real-$\omega$ curve crosses the branch cut
twice, forming a loop around the branch point (green curves), which manifests
as band backbending in the projected bandstructure plots.
It leads to a point of purely real negative BS derivative at the absorption
maximum and two points with vanishing real part of the BS derivative.

The existence of points with undefined or negative BS derivatives are
troublesome, because they could indicate unusual behavior in the tDOS.
On the other hand, the appearance of an avoided crossing between the optical 
band and a fixed resonance might cause unusual features in the tDOS, as well.
In Fig.~\ref{fig:bulk_DOS_comparison}, we compare expressions for the tDOS based
on the full inverse BS derivative (\ie $\partial k / \partial\omega$) 
evaluated on the real-$\omega$ and the real-$k$ bandstructures with our 
result derived from Eq.~\eqref{eqn:DOS_final} (note: we excluded the 
disconnected second band from the real-$\myvec k$ expression in panel a).
Indeed, the real-$k$ result incorrectly shows a divergence of the tDOS around 
the resonance as soon as the strong coupling regime is entered.
Conversely, the real-$\omega$ result incorrectly predicts a completely 
vanishing tDOS at the resonance in the case of a cusp in the real-$\omega$ 
contour and a frequency range with negative tDOS in the presence of a loop in 
the real-$\omega$ contour.
The actual tDOS derived from the bulk Green tensor does exhibit a slight wavy 
perturbation due to the presence of a material resonance, but remains 
well-behaved and does not exhibit any discontinuous behavior for any finite
damping rate $\gamma$~\cite{Barnett:1996}.
It should be noted that the different tDOS-like curves differ significantly 
even for fairly low oscillator strengths (Fig.~\ref{fig:bulk_DOS_comparison}b) 
in a homogeneous material.
This illustrates the significance of Section~\ref{sec:DOS} for the analysis of
emission dynamics in structured dispersive systems.
As an example of a slightly more complex system, we here mention Plasmon-exciton polaritons in 2D semiconductor/metal interfaces where the exciton line resembles our Lorentzian resonance, while the surface-plasmon polariton represents a strongly dispersive medium as the Lorentzian resonance is tuned toward the surface-plasmon resonance~\cite{Goncalves:2017}.

\subsection{Perturbation theory for non-degenerate bands}
\label{sec:perturb_band}

We will now derive the modification of the real-$\omega$ contour of the complex
bandstructure due to a change in the permittivity or permeability.
We focus on this contour, because it is the relevant one \eg for the 
computation of densities of states.
An analogous derivation for the real-$\myvec k$ (or any other) contour is
straight-forward.
We chose the ``energy form'' and start by introducing perturbed material 
distributions:
\begin{align}
  \eps' = & \eps + \Delta \eps, & \mu' = & \mu + \Delta \mu.
\end{align}
As a result, the expressions for $\mathcal{W}$, $\mathcal{L}$ and 
$\mathcal{P}_z$ are also modified according to:
\begin{align}
  \mathcal{W}' = \mathcal{W} + \Delta \mathcal{W} =
  \mathcal{W} + 
  \frac{1}{2} \left( \begin{array}{cc}
  \Delta \mu & 0 \\
  0 & \Delta \eps
  \end{array} \right), 
\end{align}
One property of the ``energy form'' (as well as the ``Lagrangian form'') is 
that the products $\mathcal{W} \mathcal{L}$ and $\mathcal{W} \mathcal{P}_z$ 
are not affected by this type of perturbation.
We now evaluate Eq.~\eqref{eqn:operator_perturbation_vgr} for the same 
frequency $\omega$, but using the perturbed weight function $\mathcal{W}'$:
\begin{align}
  (\psi^\ddagger, [\mathcal{L}' + \kappa \mathcal{P}_z'] \psi)^{(\mathcal{W}')}_\omega
  = & \omega (\psi^\ddagger, \psi)^{(\mathcal{W}')}_\omega.
\end{align}
In this context, $\kappa$ describes the change of the (complex) wave number in
addition to the $\omega$-dependence of $\myvec k$ found in the unperturbed band 
structure.
We now assume normalized adjoints $(\psi^\ddagger, \psi) = 1$, we substitute 
the integral definition of the bilinear form and use the identities 
$\mathcal{W}'(\mathcal{L}' + \kappa \mathcal{P}_z')
= \mathcal{W}(\mathcal{L} + \kappa \mathcal{P}_z)$ and 
$\mathcal{L} \psi = \omega \psi$ to find for the leading order in 
$\Delta \mathcal{W}$:
\begin{align}
  \kappa 
  = & \omega s_z
  (\psi^\ddagger, \, \mathcal{W}^{-1} \Delta \mathcal{W} \, \psi)
  \label{eqn:wave_number_perturbation}
  \\
  = & \frac{\omega s_z}{2}
    \int_\WSC \total^3 r \ 
    \myvec e^\ddagger \Delta \eps \myvec e
    + \myvec h^\ddagger \Delta \mu \myvec h,
\end{align}
where $s_z = (\partial k_z)/(\partial \omega)$ is the projected inverse BS 
derivative introduced earlier.

From this, we can now compute the perturbative change in the Green tensor,
Eq.~\eqref{eqn:DOS_deriv_100}.
Its integrand consists of two factors: 
A projector and the band structure derivative of the auxiliary problem
$(\partial k_z) / (\partial \lambda)$.
When computing the tDOS, the projector always becomes unity assuming 
normalized adjoint modes.
Therefore, we focus on the derivative.
The perturbation leads to a change in the wave number $k_z(\lambda)$ to a new
value $k_z'(\lambda)$:
\begin{align}
  k_z'(\lambda) = & k_z(\lambda) + \frac{\partial k_z}{\partial \lambda}
  \cdot \lambda (\psi^\ddagger, \, \mathcal{W}^{-1} \Delta \mathcal{W} \, \psi),
\end{align}
where we adapted Eq.~\eqref{eqn:wave_number_perturbation} to the auxiliary
eigenvalue problem.
Its $\lambda$-derivative is:
\begin{align}
  \frac{\partial k_z'}{\partial \lambda} = & 
  \frac{\partial k_z}{\partial \lambda} 
  \Big[ 1 + (\psi^\ddagger, \, \mathcal{W}^{-1} \Delta \mathcal{W} \, \psi)\Big] 
  \nonumber
  \\
  & \quad + \frac{\partial^2 k_z}{\partial \lambda^2}
  \cdot \lambda (\psi^\ddagger, \, \mathcal{W}^{-1} \Delta \mathcal{W} \, \psi).
  \label{eqn:perturbative_deriv}
\end{align}
This result is only valid for non-degenerate bands and away from points
where both $(\partial k_z)/(\partial \lambda)$ and 
$(\partial^2 k_z)/(\partial \lambda)^2$ diverge, \ie away from proper band
edges for example in lossless PhCs.
Assuming that the band curvature is weak, \ie 
$(\partial k_z)/(\partial \lambda) \gg \lambda (\partial^2 k_z)/(\partial \lambda)^2$, 
we can neglect the band curvature term and insert into 
Eq.~\eqref{eqn:DOS_final} to find for the first-order correction to the tDOS:
\begin{align}
  \DOS'(\omega) = \frac{4 \omega}{c^2}
  \int_{\Isofreq_z'(\omega)} 
  \!\!\!\!\!\!\!\!\!\!  \total^2 k \ 
  \Re\bigg\{ 
    \frac{ 
      1 + (\psi^\ddagger_{n\myvec k}, \, \mathcal{W}^{-1} \Delta \mathcal{W} \, \psi_{n\myvec k}) 
    }{
      (\psi^\ddagger_{n\myvec k}, \mathcal{P}_z \psi_{n\myvec k}) 
  } \bigg\}.
  \label{eqn:perturbative_tDOS}
\end{align}
From this result, we can see that in spectral ranges with predominantly real
band structure derivative, the addition of a perturbation modifies the tDOS
mostly with a term proportional to $1 + \Re\{\Delta \mathcal{W}\}$.
In contrast, the modification of the tDOS is mostly proportional to 
$\Im\{\Delta \mathcal{W}\}$ whenever the original band structure derivative is
predominantly imaginary (\ie in stop bands and band gaps).

As an example, we now assume that the perturbation of the material functions 
is given by a narrow-band resonance in the dielectric function, which implies 
$\Delta \mu = 0$.
As before we add one pole:
\begin{align}
  \Delta \eps(\myvec r, \omega) & = 
  \frac{\Theta(\myvec r)}{\omega - \widetilde \Omega},
\end{align}
where $\widetilde \Omega$ still comprises the resonance frequency and damping  
and the real-valued function $\Theta(\myvec r)$ is the spatial distribution of 
the Lorentzian perturbation.
It could be for example the product of the oscillator strength and the spatial 
distribution of color centers in a dielectric or of dye molecules in some 
solvent that infiltrates the voids of a periodic structure.
We find for the perturbative change in the wave number:
\begin{align}
  \kappa = 
  \frac{s_z \omega}{2} \cdot
  \frac{1}{\omega - \widetilde \Omega} 
  \underbrace{
    \int_\WSC \total^3 r \ \myvec e^\ddagger \Theta(\myvec r) \myvec e
  }_{2 \alpha},
\end{align}
where the real-valued parameter $\alpha$ reflects the overlap between 
individual oscillators and the electromagnetic eigenmodes.
It can be regarded as an oscillator strength density.

Assuming a narrow-band Lorentzian, we can assume the BS derivative to be 
constant across the resonance frequency $\omega = \Omega$.
Then, the real-$\omega$ contour in the spectral vicinity of the resonance 
becomes:
\begin{align}
  \myvec k'(\omega) \approx \myvec k(\Omega) - s_z \Omega + s_z \omega 
  \left(
    1 + \frac{\alpha}{\omega - \widetilde \Omega}
  \right),
\end{align}
where $\myvec k'$ and $\myvec k$ refer to wavenumbers of the perturbed and
unperturbed systems, respectively.
This equation shows the same four regimes discussed in 
Fig.~\ref{fig:coupling_cases} and Fig.~\ref{fig:bulk_DOS_comparison}.
For example, the case `c' (existence of a cusp in real-$\omega$ contour and
onset of band backbending) corresponds to the existence of a real frequency 
with vanishing BS derivative:
\begin{align}
  \frac{\partial \myvec k'}{\partial \omega}
  = s_z \left[
    1 + \frac{\alpha}{\omega - \widetilde \Omega} 
    - \frac{\alpha \omega}{(\omega - \widetilde \Omega)^2}
  \right] 
  = 0.
  \label{eqn:perturbative_cusp}
\end{align}
From the assumption that $\alpha$ is real-valued, we find that the zero occurs
at the frequency $\omega = |\widetilde \Omega| = \sqrt{\Omega^2 + \gamma^2}$.
Upon back-substituting this in Eq.~\eqref{eqn:perturbative_cusp}, we find:
\begin{align}
  \alpha = 2 (\Omega - |\widetilde \Omega|) 
  \quad \approx - \frac{\gamma^2}{\Omega} 
  \quad \text{for} \quad \gamma \ll \Omega.
\end{align}
Weaker damping (or stronger overlap) leads to case `d', stronger damping (or 
weaker overlap) leads to the cases `b' and then `a'. 
It should be noted that the BS derivative $1/s_z$ of the unperturbed 
system does not enter this relationship.
Finally, we can determine the effect of a resonant pole on the tDOS of a 
lossless periodic structure.

\subsection{Perturbation theory for parabolic band edges}
\label{sec:perturb_band_edge}

One question of particular interest is to which extent a material resonance
at the band edge of a band gap material modifies the density-of-states.
The motivation for this question is the problem that the singularities and
band gaps of photonic crystals are used to control the enhancement and 
suppression of spontaneous emission from quantum emitters.
However, usually the effect of the mere presence of resonantly polarizable
particles on the DOS is neglected, which is generally questionable in the
strong coupling limit.
In the past, the effect of non-dispersive loss on the density-of-states near a 
parabolic photonic band edge has been studied~\cite{Pedersen:2008}.

As our final example, we now present the effect of a dispersive perturbation on
the tDOS near a parabolic band edge described by the relation
\begin{align}
  k_z(\omega) = & k_0 - A \sqrt{\omega_0 - \omega},
\end{align}
where $\omega_0$ and $k_0$ are the band edge frequency and the wave number at 
which it occurs, respectively.
We assume that the unperturbed PhC is lossless (which implies the absence of
material dispersion and also that $k_0$ is real-valued) both for simplicity 
and because lossy and dispersive PhCs can be treated using the perturbation 
theory outlined in Sec.~\ref{sec:perturb_band_edge}.
The latter is not possible for a band edge in a lossless PhC, because all 
derivatives of $\myvec k$ with respect to $\omega$ diverge, \ie it constitutes 
a singularity in the real-$\omega$-contour.
Since we aim to include a perturbation with material dispersion, we again 
assume a fixed physical frequency $\omega$ and study the frequency-like 
eigenvalue $\lambda$.

\begin{figure}
  \includegraphics[width=\columnwidth]{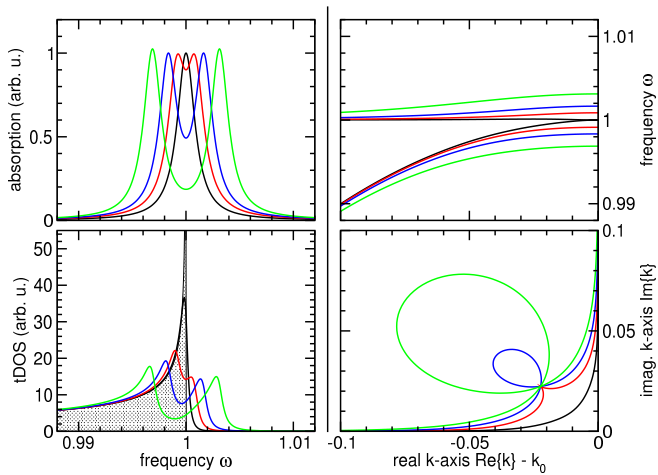}
  \caption{
    (color online)
    Comparison of the loss spectrum (top left panel), tDOS (bottom left) and 
    two projections of the complex dispersion relation (right column) of hybrid
    system composed of a two-level system strongly coupled to an isotropic band
    edge.
    The parameter varied is the oscillator strength density $\alpha$:
    $-2\times10^{-7}$ (black), $-1\times10^{-6}$ (red), $-3\times10^{-6}$ (blue)
    and $-1\times10^{-5}$ (green).
    The tDOS of the unperturbed band edge is shown as a shaded area.
    See text for further discussion.
  }
  \label{fig:band_edge_DOS}
\end{figure}

The perturbation $\Delta \mathcal{L}(\omega)$ of the Lorentzian resonance to 
the Maxwell operator $\mathcal{L}$ of the unperturbed system leads to a 
correction $\Delta$ in the eigenvalue $\lambda$ according to
\begin{align}
  [\mathcal{L} + \Delta \mathcal{L}(\omega)] \psi = & (\lambda + \Delta) \psi 
  = \lambda' \psi;
  \\
  \Delta = & \frac{(\psi^\ddagger, \Delta \mathcal{L}(\omega) \psi)_\omega}{(\psi^\ddagger, \psi)_\omega}.
\end{align}
From the parabolic dispersion relation of the perturbed system near the original
band edge $k' = A\sqrt{\omega_0 - \lambda'}$, we can therefore find the 
tDOS:
\begin{align}
  \frac{\partial k'}{\partial \lambda} 
  = & - \frac{A}{2 \sqrt{\omega_0 - \lambda - \Delta}};
  \\
  \rho_t(\omega) = & \Re\bigg\{
  \frac{2 A \omega}{\pi}
  \bigg(\omega_0 - \omega - \frac{\alpha}{\omega - \widetilde \Omega}\bigg)^{-1/2}
  \bigg\}.
\end{align}
This is the dispersive generalization of the approach chosen by Pedersen 
\emph{et al.} for the case of a PhC composed of lossy yet dispersion-free 
constituent materials~\cite{Pedersen:2008}.
An example for our result is shown in Fig.~\ref{fig:band_edge_DOS} for the
parameters $A = \omega_0 = \Omega = 1$, $\gamma = 10^{-3}$ and varying 
oscillator strength density $\alpha$.
It shows how the loss of a weakly coupled oscillator regularizes the divergent
tDOS of the ideal band edge in analogy to the non-dispersive 
case~\cite{Pedersen:2008}.
Increasing the oscillator strength, however, leads to the familiar signature
of strong coupling: a Rabi-splitting in the real-$\myvec k$ band structure 
(top right panel in Fig.~\ref{fig:band_edge_DOS}) and in the loss and 
scattering spectrum (top left panel) as well as the onset of band back-bending 
in the real-$\omega$ contour and a splitting in the tDOS, which is a bit 
smaller than the Rabi splitting.
This is of some interest in the study of the emission spectra from quantum
emitters that are strongly coupled to a lattice resonance.
The spectrum of the emission of the PhC can be expected to be the product of
the loss spectrum and the tDOS of the composite system, whose splitting is a 
bit smaller than the Rabi splitting.
As a result, the splitting found in the spectrum of photons emitted into the 
PhC is smaller than the Rabi splitting.
However, in experiments photons are detected outside the PhC, \ie the emitted
radiation must pass the PhC-air interface, which can significantly modify both
the spatial and spectral distribution of the emitted 
radiation~\cite{Schuler:2009,Frohlich:2014} and clearly is beyond the scope of 
this simple example section.

\section{Discussion and Conclusions}
\label{sec:conclusions}

In the previous sections, we have investigated the band structure in periodic
photonic systems involving material dispersion and loss based on an adjoint
modal representation.
We have focused our attention on the band structure derivative 
$(\partial \omega)/(\partial \myvec k)$ and the density of transverse photonic 
states $\rho_t(\omega)$, obtained general expressions that lend themselves very 
well to numerical evaluation and we have analytically analyzed a few simple
examples, especially highlighting the dangers of simply applying expressions
known from lossless periodic structures to cases involving dispersive 
constituent materials such as metals or narrow-band resonances. 
Our findings are in agreement with the established literature on emission 
dynamics in lossy and dispersive media~\cite{Tip:1998,Barnett:1996}, but allow
for efficient evaluation in general periodic systems.

Our main focus was the relationship between the band structure derivative 
(sometimes also referred to as complex group velocity) and the density of 
transverse optical states (tDOS).
We found that the well-known connection between the tDOS and the real part
of the group velocity still holds in case of frequency-independent material
loss, but ceases to do so in the case of dispersive materials.
The main reason for this is the fact that at each wave vector point $\myvec k$ 
inside the first Brillouin zone the dispersion relation with dispersive 
material parameters can support eigenfunction field patterns that are linearly 
dependent, \ie it supports more modes than are required for a basis of the
function space of physical modes.
The reason for this are the internal degrees of freedom of the dispersive 
material that can be treated via auxiliary differential equations leading to 
the Hamiltonian treatment by Tip~\cite{Tip:1998}.
In contrast, we eliminated the need for additional degrees of freedom by 
expanding the transverse Green tensor in an eigenvalue problem at fixed
frequency and found the tDOS in terms of a group-velocity-like quantity,
Eq.~\eqref{eqn:DOS_final}.

Throughout this study, we have assumed periodicity, which mainly includes bulk
materials, 3D photonic crystals and metamaterials with right-handed, left-handed
or hyperbolic effective material parameters.
It can be further generalized straightforwardly to systems which lack 
periodicity in any number of dimensions, provided the relevant states are 
square-integrable with respect to the non-periodic dimensions.
For such problems, the adjoint operator is still found by partial integration
as described in Appendix~\ref{appx:adjoint}, with the only difference that the
surface integral appearing from the divergence theorem vanishes for the 
non-periodic directions due to localization and for the periodic ones as
described in the Appendix.
This wider class of problems includes mostly guided modes \eg inside slab or
ridge waveguides, in PhC slabs, the coupling between excitons and structured 
metal surfaces and particle plasmon polaritons and meta-surfaces that support
bound surface states and guided lattice resonances.
It also can be applied to leaky modes provided the field pattern decays at
infinity, \eg due to some absorption in the surrounding material.
In all these cases, the adjoint operator and adjoint fields take the form of
Eqs.~(\ref{eqn:adjoint_operator_energy_nondispersive},\ref{eqn:adjoint_mode_connection_nondispersive}),
except that the wave vector is restricted to the periodic dimensions.

In our examples, we demonstrated that attempting to compute the density-of-
states of a lossy system involving material dispersion by evaluating the 
dispersion relation for real-valued wave vectors, ignoring the imaginary part
of $\omega$ and then applying the relationship between group velocity and DOS 
known from lossless problems can lead to a grossly incorrect result, especially
in the strongly coupled regime with an avoided crossing.
We further demonstrate that attempting a similar procedure by evaluating the
dispersion relation with real-valued frequencies and ignoring the imaginary
part of the wave vector equally leads to incorrect results.

\section{Acknowledgments}
N.~A.~M. is a VILLUM Investigator supported by VILLUM FONDEN (grant No. 16498). The Center for Nano Optics is financially supported by the University of Southern Denmark (SDU 2020 funding). We acknowledge discussions with M.~J.~A.~Smith, C.~M.~de~Sterke, M.~Wubs, and C.~Tserkezis.

\begin{appendix}

\section{Explicit derivation of adjoint operator and fields}
\label{appx:adjoint}
In this Appendix, we provide an explicit derivation of the relationship 
between Maxwell operator and Bloch modes and their respective adjoints with 
respect to the ``energy form''.
To this end, we reformulate $(\psi_1^\ddagger, \mathcal{L} \psi_2)$ into
$(\mathcal{L}^\ddagger \psi_1^\ddagger, \psi_2)$, starting with:
\begin{align}
  ( \psi_1^\ddagger, \mathcal{L} \psi_2 )^\mathrm{(\Energy)}
  = &
  \int_{\WSC} \total^3 r \  
  \myvec e_1^\ddagger \cdot [(\nabla + \imag \myvec k) \times \myvec h_2]
  \label{eqn:adjoint_deriv_1}
  \\
  & \quad
  \nonumber
  -
  \int_{\WSC} \total^3 r \ 
  \myvec h_1^\ddagger \cdot [(\nabla + \imag \myvec k) \times \myvec e_2].
\end{align}
We apply the identity
$\nabla \cdot (\myvec f \times \myvec g) = \myvec g \cdot 
(\nabla \times \myvec f) - \myvec f \cdot (\nabla \times \myvec g)$
to the first integral:
\begin{align}
  \nonumber
  &
  \int_{\WSC} \total^3 r \ 
  \myvec e_1^\ddagger \cdot [(\nabla + \imag \myvec k) \times \myvec h_2]
  \\
  = &
  \int_{\WSC} \total^3 r \ 
  \div (\myvec h_2 \times \myvec e_1^\ddagger)
  \\
  \nonumber
  & \quad + \int_{\WSC} \total^3 r \ \myvec h_2 \cdot 
  [(\nabla - \imag \myvec k) \times \myvec e_1^\ddagger].
\end{align}
Using Gauss's theorem, we transform the first term to an integral over the 
surface of the Wigner--Seitz cell. 
Within this integral, opposing faces cancel each other, because the functions 
$\myvec e_1^\ddagger$ and $\myvec h_2$ are periodic and the respective normal
vectors point in opposite directions.
Treating the second integral in \eqnref{eqn:adjoint_deriv_1} likewise leaves us
with:
\begin{align}
  ( \mathcal{L}^\ddagger \psi_1^\ddagger, \psi_2 )^\mathrm{(\Energy)}
  = &
  \int_{\WSC} \total^3 r \ \myvec h_2 \cdot 
  [(\nabla - \imag \myvec k) \times \myvec e_1^\ddagger]
  \\
  \nonumber
  & \quad
  - \int_{\WSC} \total^3 r \ \myvec e_2 \cdot 
  [(\nabla - \imag \myvec k) \times \myvec h_1^\ddagger].
\end{align}
Using the identity
$
(\mathcal{L}^\ddagger \psi_1^\ddagger, \psi_2)
= (\psi_1^\ddagger, \mathcal{L} \psi_2)
= \omega (\psi_1^\ddagger, \psi_2)$
we find that the adjoint modes satisfy the equation
\begin{align}
  \mu^{-1} (\nabla - \imag \myvec k) \times \myvec e^\ddagger 
  = & -\imag \omega \myvec h^\ddagger,
  \label{eqn:energy_adjoint_eqns_1}
  \\
  \eps^{-1} (\nabla - \imag \myvec k) \times \myvec h^\ddagger 
  = & \imag \omega \myvec e^\ddagger,
  \label{eqn:energy_adjoint_eqns_2}
  \\
  \Rightarrow \quad \psi^\ddagger(\myvec r, \myvec k, \omega, \eps, \mu) 
  = & \psi(\myvec r, -\myvec k, -\omega, \eps, \mu).
\end{align}
In the lossless case, this reduces to the familiar relationship
$ \psi^\ddagger(\myvec r, \myvec k)
= \psi^\ast(\myvec r, \myvec k)$.

Our expression for the (electric) adjoint modes obtained from the 
``energy form'' (the natural generalization of the standard scalar product in
lossless electromagnetism) is in conflict with the definition of adjoint modes
employed in Ref.~\cite{Chen:2010}, which incidentally is obtained 
when repeating the same steps starting with the ``Lagrangian form'':
\begin{align}
  \psi^\ddagger(\myvec r, \myvec k, \omega, \eps, \mu) 
  = & \psi(\myvec r, -\myvec k, \omega, \eps, \mu).
\end{align}
The magnetic adjoint modes are not explicitly given in Ref.~\cite{Chen:2010},
but differ by a minus sign from our ``Lagrangian adjoint'', which explains
the overall minus sign in the numerator of Eq.~\eqref{eqn:vgr_lagrangian}.

\section{Multivaluedness due to material resonances}
\label{appx:multivalue}

Here, we show that the addition of a pole in the material response of a 
periodic optical system results in the appearance of additional bands and
therefore causes the eigenmodes for a given $\myvec k$ to become linearly 
dependent.
We assume that the material properties $\eps$ and/or $\mu$ are nondispersive 
except for a resonance at some complex frequency $\Omega$.
Then, we can decompose the total wave operator:
\begin{align}
  \mathcal{L}(\myvec k, \omega) = &
  \mathcal{L}_0(\myvec k) + \frac{\mathcal{L}'(\myvec k)}{\omega - \Omega}.
\end{align}
Now we approximate the true operators as a sequence of discrete operators by
representing them in some spatial function basis (\eg a plane-wave basis) with 
increasing number $N$ of functions.
These discrete operators can be expressed as matrices of dimension $N$.
The eigenfrequencies of the non-dispersive part satisfy
\begin{align}
  \det[\mathcal{L}_0(\myvec k) - \omega] = P_0(\myvec k, \omega) = 0,
\end{align}
where $P_0(\myvec k, \omega)$ is a polynomial of degree $N$ in both $\omega$
and $\myvec k$.
As a result, the nondispersive eigenvalue problem has $N$ solutions, which 
form a basis for the complete base function space as long as $\mathcal{L}_0$ 
is not defective (guaranteed for Hermitian $\mathcal{L}_0$).

In contrast, the eigenfrequencies for the dispersive problem satisfy:
\begin{align}
  0 = &
  \det[\mathcal{L}_0(\myvec k) + 
  (\omega - \Omega)^{-1} \mathcal{L}'(\myvec k, \omega) - \omega]
  \\
  = &
  \frac{
    \det[(\omega - \Omega) \mathcal{L}_0(\myvec k) + 
    \mathcal{L}'(\myvec k) - \omega (\omega - \Omega)]
  }{ (\omega - \Omega)^N }
  \\
  = & \frac{1}{(\omega - \Omega)^N} P'(\myvec k, \omega).
\end{align}
The zeros of this are given by the zeros of $P'(\myvec k, \omega)$, which is a 
polynomial of degree $N$ in $\myvec k$ and of $2N$ in $\omega$.
As a result, the number of eigenfrequencies will be in general greater than the
matrix rank, \ie additional frequency bands appear due to the resonance 
and the set of physical eigenstates is in general linearly dependent.
In particular, the set of physical eigenstates no longer forms a basis for any
function space.

\section{Integration of the diagonal Green tensor elements in 3D}
\label{appx:integration}

\subsection{Contribution from hairpin sections for 
$\myvec r' \rightarrow \myvec r$}
\label{appx:integration_hairpin}

The integration paths connecting the individual band integrals (shown as red
hairpin-shaped paths in Fig.~\ref{fig:integration}e) come in pairs of 
$k_z$ and $-k_z^\ast$.
If the integrand in Eq.~\eqref{eqn:green_expanded_100} is invariant under the 
transformation $k_z \rightarrow -k_z$, every such pair of hairpin sections is 
equivalent to one bone-shaped integral around a complete branch cut, \ie a 
closed integral of an analytic function, which is zero, because the contour 
does not include any poles.
In this case, connecting the integration segments around the branch cuts does
not change the result of the integral.
This is the case in the limit $\myvec r' \rightarrow \myvec r$ assuming that
$\mathcal{W}(\myvec r')$ is a continuous function, an assumption that we can
live with in physical problems and that is strictly correct for the plane-wave 
expansion or related spectral expansions.

To see the equivalence, we choose the ``Lagrangian form'', whose adjoint modes
satisfy the relationship~\cite{Chen:2010}:
\begin{align}
  \Phi^\ddagger(\myvec r, t, \myvec k, \lambda) = 
  \Phi(\myvec r, t, -\myvec k, \lambda).
\end{align}
This also implies the existence of a pole at $\lambda_{n -\myvec k \omega}$.
Furthermore, in the presence of time reversal the tensor product 
$[\Phi_{n\myvec k \omega}(\myvec r) \otimes \Phi^\ddagger_{n\myvec k \omega}(\myvec r')]$ is symmetric under matrix transposition.
Therefore, the integrand satisfies the identity:
\begin{align}
  & \frac{
    [\Phi_{n\myvec k \omega}(\myvec r) \otimes
    \Phi^\ddagger_{n\myvec k \omega}(\myvec r')]
  \mathcal{W}(\omega, \myvec r')}{\lambda - \lambda_{n \myvec k \omega}}
  \\
  = &
  \frac{
    [\Phi_{n (-\myvec k) \omega}(\myvec r) \otimes
    \Phi^\ddagger_{n (-\myvec k) \omega}(\myvec r')]
  \mathcal{W}(\omega, \myvec r)}{\lambda - \lambda_{n (-\myvec k) \omega}}.
\end{align}
Note that the argument of $\mathcal{W}$ changed from $\myvec r'$ to $\myvec r$.
Thus they are the same in the limit $\myvec r' \rightarrow \myvec r$ provided
that $\mathcal{W}$ is continuous.
It can be easily seen that the integrand does not depend on whether one chooses
the ``Lagrangian form'' or the ``energy form'', so the results apply to the 
latter as well.

\subsection{Extended zone integration in 3D and isocontours} 
\label{appx:integration_3d}

In Sec.~\ref{sec:DOS}, we outlined how the collocal elements of the transverse 
Green function can be found in a 1D periodic system.
This can be generalized to 2D and 3D periodicity, by choosing one reciprocal
lattice primitive $\myvec G_1$ and treat $\myvec k$-space along this direction
as described for the 1D problem, \ie one allows for a non-real contribution to 
the total wave vector in this direction and one moves on to an extended zone 
scheme along this direction.
For simplicity and without loss of generality, we refer to this as the 
$z$-direction.
In contrast, the contributions $\myvec k_\perp$ to the wave vector from the 
other reciprocal lattice primitives $\myvec G_2, \myvec G_3$ are kept 
real-valued and in the reduced zone scheme. 
We call this the real wave vector offset.
The resulting integration zone in $\myvec k$-space is the tensor product of a
parallelogram $\Omega_\perp$ spanned by the reciprocal lattice primitives 
$\myvec G_2$ and $\myvec G_3$ and the full complex plane oriented along the 
reciprocal lattice primitive $\myvec G_3$.
Within this domain, we integrate over the real-$\lambda$-contour.
For all real wave vector offsets and all unique bands therein, we integrate 
along the complex $\myvec G_1$-direction as described in Sec.~\ref{sec:DOS}, 
\ie encircling a branch point at each lattice resonance except for the 
$\Gamma$-point.
We end up with an integral of the form
\begin{align}
  \sum_n
  \int_{\Omega_\perp} \!\!\! \total^2 k_\perp
  \int_{-\infty}^\infty \!\!\!\! \total k_z
  \int_{-\infty}^\infty \!\!\!\! \total \kappa
  \ \delta(\lambda - \lambda_{n (\myvec k \myvec G_1) \omega}) \ 
  f_\omega(\myvec k, \lambda),
  \label{eqn:3d_integration_proper}
\end{align}
where $\myvec k = \myvec k_\perp + (k_z + \imag \kappa) \myvec G_1$, $n$ labels
the unique band indices and $f_\omega$ is the function to be integrated, in 
our case the projector onto the respective eigenmode.
As a simplified notation for this, we introduce the symbol 
$\Isofreq_{z, \omega}(\lambda)$, which is to be understood as the 
$\lambda$-isocontour in the $\myvec k$-space integration domain.
The expression
\begin{align}
  \int_{\Isofreq_{z,\omega}(\lambda)} 
  \!  \!  \!  \!
  \total^2 k \ \frac{\partial k_z}{\partial \lambda} \ 
  f_\omega(\myvec k, \lambda),
\end{align}
is understood to be a simplified notation for the integral shown in 
Eqn.~\eqref{eqn:3d_integration_proper}.

\end{appendix}


%

\end{document}